\documentclass[sigconf, screen, table]{acmart}

\usepackage{csquotes}
\usepackage{graphicx}
\usepackage{lipsum}
\usepackage{dirtytalk}
\usepackage{soul} 
\usepackage{xcolor}
\usepackage{soul}
\sethlcolor{yellow!20}

\usepackage{soul} 
\renewcommand{\hl}[1]{#1} 

\AtBeginDocument{%
  }

\setcopyright{acmlicensed}
\copyrightyear{2026}
\acmYear{2026}
\acmDOI{XXXXXXX.XXXXXXX}
\acmConference[FAccT'26]{Proceedings of the 9th ACM Conference on Fairness, Accountability, and Transparency}{June 25--28, 2026}{Montreal, Canada}
\acmISBN{978-1-4503-XXXX-X/2018/06}

\begin{document}

\title[How Researchers Navigate Accountability, Transparency, and Trust When Using AI Tools]{How Researchers Navigate Accountability, Transparency, and Trust When Using AI Tools in Early-Stage Research: A Think-Aloud Study}

\author{Sanjana Gautam}
\authornote{Work performed as a Postdoctoral Fellow at UT Austin prior to joining Microsoft.}
\affiliation{%
  \institution{Microsoft}
  \city{Redmond}
  \state{WA}
  \country{USA}
}

\author{Houjiang Liu}
\affiliation{%
  \institution{The University of Texas at Austin}
  \city{Austin}
  \state{Texas}
  \country{USA}
}

\author{Yujin Choi}
\affiliation{%
  \institution{The University of Texas at Austin}
  \city{Austin}
  \state{Texas}
  \country{USA}
}

\author{Matthew Lease}
\affiliation{%
  \institution{The University of Texas at Austin}
  \city{Austin}
  \state{Texas}
  \country{USA}
}

\renewcommand{\shortauthors}{Gautam et al.}

\begin{abstract}

In the early stages of scientific research, researchers rely on core scholarly judgments to identify relevant literature, assess credible evidence, and determine which directions merit pursuit. As AI tools become increasingly integrated into these early-stage workflows, the scholarly judgments that were once transparent and attributable to individual researchers become obscured, raising critical Responsible AI (RAI) concerns around accountability, transparency, and trust. Yet how these three dimensions manifest in real-time, in-situ scholarly practice remains largely unexplored. To address this gap, we conducted a think-aloud study with 15 researchers to examine how they used AI tools powered by large language models (LLMs) across early-stage research tasks, including literature exploration, synthesis, and research ideation. Our key findings address the tripartite constructs of accountability, transparency, and trust. First, the confident tone of AI outputs misrepresents epistemic uncertainty, making it more difficult for researchers (who are ultimately accountable) to identify which outputs require the greatest scrutiny. Second, opaque retrieval and content construction make provenance difficult to establish for transparency. Third, trust in AI is fragile, context-dependent, and easily eroded. In response, participant researchers were seen to develop compensatory strategies to restore scholarly judgment under uncertainty. Overall, our findings serve to contextualize AI-mediated research as a RAI problem grounded in lived researcher experience and motivate attention to deliberate AI integration that preserves accountability, supports transparency, and fosters informed trust.

\end{abstract}



\received{13 January 2025}

\maketitle

\section{Introduction}

Traditional scientific exploration (early-stage)is rooted in theoretical derivations, experimental observations, and numerical simulations \cite{klahr1999studies, granjeiro2025future}. Researchers rely on core scholarly judgments to guide their work, such as identifying which previous studies are relevant, assessing what qualifies as reliable evidence, and deciding which lines of investigation merit further pursuit \citep{Booth2009-yh}. These judgments have traditionally been the responsibility of human researchers, shaped by formal training, collaborative conversations and years of experience. However, because AI tools are now increasingly integrated with literature exploration \citep{spillias2024human}, idea generation \citep{pu2025ideasynth, liu2025personaflow}, study planning \citep{Feng2026-eb}, and other research activities \citep{al2024impact, chen2025ai4research}, this shift makes scholarly judgments that were once transparent and clearly attributable to individual researchers increasingly obscured \citep{wagman2025generative, carobene2024rising, Binz2025-bz}. Notably, the risk is not only that AI tools may produce poorly grounded outputs, but also that researchers may lose insight into, and control over, the judgments that ultimately shape their work.

The integration of AI into scientific research fundamentally reshapes how epistemic responsibility is distributed, negotiated, and justified within scholarly communities. While the affordances of AI offer significant acceleration, they also surface critical questions about how scientists interpret and enact accountability, transparency, and trust in practice \cite{granjeiro2025future}. As AI reshapes how scientific contributions are produced and justified, three responsible AI (RAI) dimensions become central: \textbf{accountability}, \textbf{transparency}, and \textbf{trust}. These dimensions are widely discussed in scholarship on the integration of LLMs into scientific research \citep{Binz2025-bz, douglas2025researchers}. Accountability remains critical because researchers retain full responsibility for scientific claims, even when those claims are derived from, refined by, or generated with LLMs. Transparency is likewise a major concern, as LLM use can obscure the provenance and interpretation of evidence, making research claims more difficult to assess and verify. Trust, in turn, is not an inherent attribute of the technology but a relational and situated concept, which is task-specific and contingent on how researchers actively cultivate and maintain responsible AI use in practice.

Prior work has documented how researchers perceive and use AI tools through surveys and interviews \cite{liang2024mapping}. Large-scale survey studies, for instance, mapped usage patterns and perceptions across disciplinary and demographic groups \cite{liao2024llms, arroyo2025generative, schroeder2024large}. Qualitative studies (predominantly semi-structured interviews) also provided richer, narrative-level insight into researchers' motivations, concerns, and normative negotiations around AI integration \cite{morris2023scientists, kapania2025m}. While these studies provide valuable insights, the concrete empirical tensions surrounding accountability, transparency, and trust that arise in real-time and in situ scholarly judgments shaped by AI have largely yet to be examined. In this work, we focus on two key research questions: 
\vspace{-0.5em}
\begin{itemize}
    \item \textbf{RQ1} What accountability, transparency, and trust concerns emerge when researchers use LLM-based tools during early-stage research workflows?
    \item \textbf{RQ2} What compensatory strategies do researchers develop to manage accountability, transparency, and trust concerns when using LLM-based tools during early-stage research workflows?
\end{itemize} 
\vspace{-0.5em}

To address these questions, we employed a think-aloud study with 15 researchers to examine how they used LLM-based tools on early-stage research workflows, including \textit{literature exploration, synthesis,} and \textit{research ideation}. Think-aloud study can capture scholarly judgment as it happens in real-time \citep{fonteyn1993description, charters2003use, fan2020practices}. Focusing on early-stage research workflows, where scholarly judgment is first exercised and AI mediation is less visible, allows us to identify which accountability, transparency, and trust concerns emerge and how researchers proactively respond to them in the moment.Overall, our study provides empirical evidence of how tensions around accountability, transparency, and trust emerge in real-time scholarly judgment during early-stage research workflows. We contextualize RAI problems in AI-mediated research within the lived experiences of researchers, rather than abstract ethical prescriptions. By highlighting the compensatory strategies researchers naturally develop, we show that AI-mediated research often adds cognitive burden instead of reducing it. These findings raise attention to deliberate AI integration that preserves accountability, supports transparency, and fosters informed trust.

\section{Related Work}

We review two bodies of work to situate our study. First, we examine how AI's role in research workflows has shifted from a cognitive scaffold to active participant in scholarly judgment (Section \ref{subsec:AI role}). We then review how RAI scholarship frames accountability, transparency, and trust in AI-mediated research contexts, highlighting the empirical gap our study addresses (Section \ref{subsec:RAI in AI4Science}).

\subsection{The Shifting Role of AI in Research Workflows} \label{subsec:AI role}

Research is a complex, multi-stage, and collaborative endeavor, spanning activities such as literature search, knowledge synthesis, idea generation, data analysis, and academic writing. A substantial body of human-computer interaction (HCI) research has examined how researchers engage in these activities and how computational tools can reduce cognitive overhead, support sensemaking, and scaffold the collaborative aspects of research work \cite{yahagi2025paperwave, oyelude2024artificial, khalifa2024using, razack2021artificial, pinzolits2024ai, nielbo2024quantitative, shen2022towards}.

Two trajectories of computational support have emerged, reflecting fundamentally different assumptions about the role of AI in research. The first augments researchers' existing workflows, keeping human researchers central to the epistemic process. The second relies on generative capabilities to perform the epistemic work itself, reducing researchers to evaluators. A prominent example of the first is the \textit{Semantic Reader Project} \citep{Lo2024-qe}, a suite of academic tools designed to reduce the cognitive overhead of managing large bodies of literature and enhancing researchers' sensemaking. Representative tools include \textit{PaperWeaver} \citep{yahagi2025paperwave}, which recommends newly published papers based on a researcher's existing collection; \textit{Threddy} \citep{Kang2022-we} and \textit{Synergi} \citep{Kang2023-yh}, which enable researchers to clip and organize passages from multiple papers into hierarchical threads. In these systems, AI functions as a cognitive scaffold: it reduces overhead and surfaces relevant material, while leaving the scholarly judgments themselves to the researcher.

The second trajectory goes further, using generative AI to perform scholarly judgment directly. Tools such as \textit{DiscipLink} \cite{zheng2024disciplink} support interdisciplinary literature integration by generating field-specific questions and surfacing thematic connections across disciplines. Similarly, \textit{IdeaSynth} \cite{pu2025ideasynth} integrates AI generation into the research ideation workflow: while human researchers can guide the process, the system acts the primary driver — synthesizing literature and generating research questions. Expanding early-stage research workflows to encompass the full research process reveals a fully automated end of this trajectory. Systems such as \textit{AI Scientist} \citep{lu2024ai} and \textit{Agent Laboratory} \citep{Schmidgall2025-ff} remove the researcher from the generative loop entirely, autonomously conducting literature review, hypothesis generation, experimentation, and writeup.

This shifted trajectory reveals that computational systems, like AI, are increasingly designed not merely as cognitive scaffolds that amplify researchers' existing capabilities, but as epistemic participants that actively influence what researchers attend to, how they frame problems, and which directions they pursue. It is precisely this shift -- from tool to co-constructor of the research process -- that makes understanding how researchers negotiate, appropriate, and resist these systems so consequential, motivating the empirical focus of our study.

\subsection{Contextualizing Responsible AI Problems in AI-Mediated Research} \label{subsec:RAI in AI4Science}

The shift from cognitive scaffold to active participant in scholarly judgment (as described in Section \ref{subsec:AI role}) transforms accountability, transparency, and trust from abstract ethical concerns into practical challenges that researchers must navigate in their everyday research workflows \cite{Wang2023-hl}.

The integration of AI into the scientific practice is not merely a technical challenge but a significant Responsible AI (RAI) problem. RAI scholarship has developed definitions of these three constructs that are directly relevant for understanding what is at stake in AI-mediated research:
\begin{enumerate}
    \item \textbf{Accountability} refers to the socio-technical arrangements through which specific actors are obligated to explain, justify, and bear responsibility for the impacts of a system to identifiable forums, with the possibility of oversight, correction, and sanction \cite{cooper2022accountability, gansky2022counterfacctual, laufer2022four}. Within AI-mediated research, accountability is the condition in which identifiable actors can be required to explain and justify system behavior to appropriate forums and face consequences for its impacts. 
    \item \textbf{Transparency} refers to the provision of information about socio-technical systems in forms that are accessible, interpretable, and meaningful to relevant stakeholders, such that it supports understanding, scrutiny, and the possibility of oversight rather than mere disclosure \cite{cooper2022accountability, gansky2022counterfacctual, laufer2022four, corbett2023interrogating}. Transparency in AI-mediated research represents the meaningful visibility of how AI systems source, process, and synthesize scientific information in ways that allow researchers to assess credibility, reproduce reasoning, and justify scholarly decisions.
    \item \textbf{Trust} refers to a user’s willingness to rely on an AI system under conditions of uncertainty and vulnerability, grounded in a justified belief that the system will reliably uphold relevant goals, expectations, or normative contracts \cite{gautam2025towards, manzini2024should, pareek2024trust, ferrario2022explainability, mehrotra2024integrity}. Trust in AI-supported research is a researcher’s justified willingness to rely on an AI system for scholarly tasks despite uncertainty, based on the belief that the system will uphold norms of accuracy, relevance, and integrity.
\end{enumerate}

Together, these three constructs capture the RAI challenges that arise at two levels in AI-mediated research. At the modeling level, risks include the embedding of bias in training data \citep{Bender2021-zi} and the use of opaque, black-box systems with limited interpretability \citep{Rudin2019-ee}, both of which directly undermine transparency and complicate accountability. At the adoption level, risks concern how researchers negotiate stakeholder values \citep{Deshpande2022-sm, Jakesch2022-bk}, enact mechanisms of accountability, and develop working trust with AI systems in practice \citep{Liao2022-jm}. While modeling-level risks have received considerable attention in the RAI literature, adoption-level risks, and in particular how accountability, transparency, and trust concerns manifest in the real-time scholarly judgments of active researchers, remain largely unexamined empirically. It is this gap that our study seeks to address.

\section{Methodology} \label{sec:methods}

We conducted a think-aloud study to investigate how scientists engage with LLM-based tools during the early stages of the research pipeline, ranging from breadth and depth research exploration and synthesis. Our goal was to examine opportunities and challenges related to user control, trust, and task relevance in these real-world research practices. As discussed above, the early stages of research represent the moments at which scholarly judgment is simultaneously most consequential and least visible. We thus focused on these early stages rather than later stages of the research pipeline such as development \cite{o2025scientists}, evaluation \cite{shi2023understanding, yan2024human}, and writing \cite{wen2024overleafcopilot}. In the following sections, we report our participant recruitment (Section \ref{subsec:participants}), and study protocol (Section \ref{subsec:protocol}). Our study was reviewed and approved by our organization's Institutional Review Board (IRB).

\subsection{Participants} \label{subsec:participants}

We recruited participants using criterion-based sampling through faculty mailing lists, personal contacts, online channels, and conference posters. Eligible participants were required to be over 18 years of age and actively engaged in research at any level. Interested individuals completed a Qualtrics recruitment survey to indicate eligibility and availability. Those selected were subsequently contacted to schedule the study sessions. 

In total, 15 participants (Table \ref{tab:participants}) took part in think-aloud interview sessions. They completed the study on their own computers, with each session conducted remotely and recorded via Zoom. Sessions lasted approximately 60–90 minutes and followed a structured format: a pre-task questionnaire, the think-aloud tasks, and a post-task questionnaire. We describe these details in Section \ref{subsec:protocol}. 

The sample consisted of eleven doctoral students, two postdoctoral researchers, one faculty member, and one industry researcher. Among the doctoral students, the majority were advanced PhD candidates. This distinction is notable, as many participants had begun their doctoral studies before the widespread availability of AI tools, and thus had already developed proficiency in conducting research without them. Participants represented \textit{a range of disciplinary backgrounds but shared familiarity with core academic workflows}, including literature review, writing, and the use of digital research tools.

\begin{table*}[]
\begin{tabular}{|l|l|l|l|}
\hline
\rowcolor[HTML]{C0C0C0} 
\textbf{Participant}       & \textbf{Gender}          & \textbf{Research Experience}                    & \textbf{Area of Research}                          \\ \hline
\rowcolor[HTML]{FFFFFF} 
{\color[HTML]{434343} P1}  & {\color[HTML]{434343} M} & {\color[HTML]{434343} Industry}                 & {\color[HTML]{434343} Steel Industry}              \\ \hline
\rowcolor[HTML]{F6F8F9} 
{\color[HTML]{434343} P2}  & {\color[HTML]{434343} F} & {\color[HTML]{434343} Researcher}               & {\color[HTML]{434343} Astrophysics}                \\ \hline
\rowcolor[HTML]{FFFFFF} 
{\color[HTML]{434343} P3}  & {\color[HTML]{434343} F} & {\color[HTML]{434343} PhD Student}              & {\color[HTML]{434343} AI Journalism}               \\ \hline
\rowcolor[HTML]{F6F8F9} 
{\color[HTML]{434343} P4}  & {\color[HTML]{434343} F} & {\color[HTML]{434343} PhD Student}              & {\color[HTML]{434343} VR \& Mental Health}         \\ \hline
\rowcolor[HTML]{FFFFFF} 
{\color[HTML]{434343} P5}  & {\color[HTML]{434343} F} & {\color[HTML]{434343} Associate Professor}      & {\color[HTML]{434343} Misinformation Studies}      \\ \hline
\rowcolor[HTML]{F6F8F9} 
{\color[HTML]{434343} P6}  & {\color[HTML]{434343} M} & {\color[HTML]{434343} Teaching Associate}       & {\color[HTML]{434343} AI \& Religion}              \\ \hline
\rowcolor[HTML]{FFFFFF} 
{\color[HTML]{434343} P7}  & {\color[HTML]{434343} F} & {\color[HTML]{434343} PhD Student}              & {\color[HTML]{434343} Neuroscience}                \\ \hline
\rowcolor[HTML]{F6F8F9} 
{\color[HTML]{434343} P8}  & {\color[HTML]{434343} M} & {\color[HTML]{434343} Post-doctoral Researcher} & {\color[HTML]{434343} Biomedical Imaging}          \\ \hline
\rowcolor[HTML]{FFFFFF} 
{\color[HTML]{434343} P9}  & {\color[HTML]{434343} F} & {\color[HTML]{434343} PhD Student}              & {\color[HTML]{434343} Industrial Labor History}    \\ \hline
\rowcolor[HTML]{F6F8F9} 
{\color[HTML]{434343} P10} & {\color[HTML]{434343} M} & {\color[HTML]{434343} PhD Student}              & {\color[HTML]{434343} Climate Science}             \\ \hline
\rowcolor[HTML]{FFFFFF} 
{\color[HTML]{434343} P11} & {\color[HTML]{434343} M} & {\color[HTML]{434343} PhD Student}              & {\color[HTML]{434343} Autonomous Driving}          \\ \hline
\rowcolor[HTML]{F6F8F9} 
{\color[HTML]{434343} P12} & {\color[HTML]{434343} M} & {\color[HTML]{434343} PhD Student}              & {\color[HTML]{434343} Library Sciences}            \\ \hline
\rowcolor[HTML]{FFFFFF} 
{\color[HTML]{434343} P13} & {\color[HTML]{434343} F} & {\color[HTML]{434343} PhD Student}              & {\color[HTML]{434343} Technology for Older Adults} \\ \hline
\rowcolor[HTML]{F6F8F9} 
{\color[HTML]{434343} P14} & {\color[HTML]{434343} F} & {\color[HTML]{434343} PhD Student}              & {\color[HTML]{434343} Education Technology}        \\ \hline
\rowcolor[HTML]{FFFFFF} 
{\color[HTML]{434343} P15} & {\color[HTML]{434343} F} & {\color[HTML]{434343} PhD Student}              & {\color[HTML]{434343} Sociotechnical Design}       \\ \hline
\end{tabular}
\vspace{0.1cm}
\caption{Participant Demographics}
\label{tab:participants}
\end{table*}

\subsection{Study Protocol} \label{subsec:protocol}

\subsubsection{Choice of Tools} \label{subsubsec:tools}

For the purposes of this study, we selected two commercially available tools that together represent distinct but complementary epistemic modalities present across the broader landscape of AI-powered research tools. To inform our selection, we conducted a structured audit of tools spanning multiple stages of the research pipeline (commercial). Rather than aiming for exhaustive representativeness across all available systems, our selection was guided by two criteria: accessibility and epistemic diversity. We deliberately prioritized commercial tools over academic prototypes, as commercial systems reflect the conditions under which most researchers currently encounter and adopt AI in practice. This can be attributed to them being actively maintained, publicly accessible, and increasingly integrated into everyday research workflows in ways that one-off academic systems often are not. \textbf{Research Rabbit} was selected as a representative of graph-based literature discovery tools, which surface relevant papers through citation network exploration and semantic similarity rather than keyword matching alone. \textbf{ElicitAI} was selected as a representative of generative, literature-grounded synthesis tools, which use LLMs to extract, summarize, and reason across collections of papers in response to natural language queries. Together, these two tools instantiate the two dominant epistemic modalities currently shaping AI-assisted early-stage research: one oriented toward exploration and network traversal, the other toward synthesis and generative elaboration. Note that our study does not aim to evaluate these specific systems, but to observe how researchers reason, negotiate, and make judgments under realistic contemporary conditions of AI-mediated research work. To preserve behavioral logs, a single sign-on profile was created for each tool, and a lowest-tier subscription plan for Elicit was maintained to support the study's recruitment timeline across multiple participants per month.

\subsubsection{Pre-task Questionnaire} \label{subsubsec:pre-task questionnaire}

The study began with a pre-task questionnaire designed to capture participant background knowledge, research practices, and prior experiences with AI tools. After providing informed consent, participants completed a survey that assessed: (i) their involvement in research, experience with literature review, and familiarity with standard academic workflows; (ii) their prior use of digital research tools, including tasks such as summarizing papers, generating ideas, writing and editing academic text, and refining research questions; (iii) their perceptions of the learning curve, integration of LLMs into research workflows, confidence in using LLMs, and interface usability; and (iv) specific tools they had used, challenges encountered, and research tasks they would not trust LLMs to perform.

\subsubsection{Think-aloud Protocol} \label{subsubsec:protocols}

Think-aloud protocols (TAPs) are research methods where participants verbalize their thoughts while completing tasks, providing insights into cognitive processes and decision-making \cite{fonteyn1993description, charters2003use}. TAPs consist of two components: the think-aloud interview for data collection and protocol analysis for analyzing verbal data \cite{hu2017using, charters2003use}. These methods are widely used across disciplines, from UX research where practitioners employ them in both controlled lab studies and remote usability testing \cite{fan2020practices}, to healthcare settings for investigating clinical decision-making \cite{lundgren2010think}. Before introducing the tasks, we define \textbf{a Thread} as a potential research direction or idea identified for further exploration. Below, we describe in detail the specific tasks within the think-aloud protocol:

\begin{enumerate}
    \item \textbf{Task 1 (Exploration in a New Domain)}: Participants began research on a new or familiar domain using their \textbf{preferred LLM-based tool}, exploring a seed paper and generating initial insights.
    \item \textbf{Task 2 (Thread Identification and Synthesis (Research Rabbit))}: Using a seed paper, participants identified 2–4 key research threads and created an outline for a potential literature review. Researchers observed tool usage for information extraction, organization, and synthesis.
    \item \textbf{Task 3 (In-Depth Thread Exploration (Elicit AI))}: Participants expanded upon a selected thread by finding additional papers via LLM-based search and scholarly databases. They brainstormed research questions and integrated cross-domain literature, allowing researchers to observe how tools bridge knowledge gaps.
    \item \textbf{Task 4 (Evaluation of AI-Generated Summaries and Outlines (Elicit AI))}: Participants assessed the quality and comprehensiveness of AI-generated summaries or outlines, refining or expanding upon outputs where necessary.
    \item \textbf{Task 5 (Reflection on Personal Workflows)}: Participants participated in semi-structured interviews to apply insights from previous tasks to their ongoing research projects, discussing tool usefulness, challenges, and potential integration strategies.
\end{enumerate}

Each task was designed to approximate real-world research scenarios, enabling participants to engage with the tools in ways that closely mirror their intended use in practice. Participants began with seed papers situated adjacent to their own research domains and, building on these, explored related literature in familiar areas. This process culminated in the formulation of potential research questions for a new study.

\subsubsection{Post-task Questionnaire} \label{subsubsec:post-task questionnaire}

Participants completed a post-task questionnaire intended to capture in-depth reflections on their task experiences, perceptions of potential future use, and evaluations of the tool’s impact on their work efficiency. The questionnaire specifically probed changes in participants’ strategies, insights gained from interacting with the tool, and assessments of the AI-generated outputs. 

\subsubsection{Data Analysis}\label{subsubsec:datanal}

In this study, we first analyzed pre-survey data to establish participant baseline use of AI tools and obtain contextual background. We then employed inductive coding to generate preliminary themes from the think-aloud interviews. Our analytic process followed a two-step approach: an initial inductive analysis, followed by iterative refinement across additional data until thematic saturation was achieved \cite{ando2014achieving}. Within thematic analysis, inductive coding offers a systematic means of capturing emergent patterns, and when applied to think-aloud protocols, it supports rigor and validation \cite{fonteyn1993description}. The think-aloud method, in turn, provides rich verbal accounts of participants’ reasoning during problem-solving tasks, yielding valuable insights into their cognitive processes \cite{ando2014achieving}.

\section{Thematic Findings} \label{sec:Results}

Our initial open coding surfaced 23 sub-themes reflecting a wide range of practices, concerns, and adaptation strategies. Through iterative axial coding and theoretical grouping \cite{vollstedt2019introduction}, we narrowed these to central themes that consistently recurred across participants and tasks, and that aligned with core RAI principles: accountability, transparency, and trust (summarized in Table \ref{tab:findings}). 

Within each dimension, we first report themes that reflect tensions and concerns around responsibility, verification, and epistemic authority that emerged as researchers interacted with AI tools in real time. We then report how researchers actively negotiate uncertainty by developing compensatory strategies. These strategies reveal that AI adoption is not a passive substitution of human judgment, but a cognitively demanding process in which researchers reconfigure their workflows to maintain control, protect their reputation, and preserve disciplinary standards.

\begin{table*}[ht]
\centering
\begin{tabular}{@{}p{6cm}p{6cm}@{}}
\toprule
\textbf{Tensions \& Concerns (RQ1)} & \textbf{Compensatory Strategies (RQ2)} \\
\midrule

\multicolumn{2}{@{}l}{\textbf{Accountability}} \\
Mismatch between assertive AI outputs and uncertain scholarly judgment. \newline
Platform-level opacity around data governance.
&
Human retains ultimate scholarly judgment. \newline
Restrict AI to organizational and peripheral tasks.
\\
\midrule

\multicolumn{2}{@{}l}{\textbf{Transparency}} \\
Blackbox opacity across retrieval and narrative construction. \newline
Hallucinations as transparency failures.
&
Social credibility heuristics. \newline
Redundant manual verification. \newline
Constrained prompting.
\\
\midrule

\multicolumn{2}{@{}l}{\textbf{Trust}} \\
A fragile, context-dependent, and continuously re-established process. \newline
Output homogenization undermines perceived intellectual depth.
&
Progressive trust-building through logical auditing. \newline
Hedging AI use to low-stakes tasks.
\\
\bottomrule
\end{tabular}
\vspace{0.1cm}
\caption{Summary of thematic findings}
\label{tab:findings}
\end{table*}

\subsection{Accountability}
Accountability concerns emerged consistently across participants, centering on two interrelated tensions: the mismatch between the assertive nature of AI outputs and the inherent uncertainty of scholarly judgment, and the opacity of how AI platforms manage researcher data. In response, participants developed strategies to retain ultimate scholarly judgment by restricting AI use to peripheral and organizational tasks rather than core epistemic work.

\subsubsection{Mismatch between assertive AI outputs and uncertain scholarly judgment} 

A fundamental tension participants identified (7 of 15) was the decoupling of AI confidence from scholarly validity. They found that AI tools consistently generated fluent, assertive outputs regardless of whether a clear or authoritative answer existed, creating a systematic mismatch with the uncertainty and ambiguity that characterizes genuine scientific inquiry. 

This underscores a mismatch of critical scientific inquiry: while scientific questions are inherently uncertain and open to interpretation, the models fail to capture these epistemic nuances. 

As P7 noted:
\begin{quote}
    ``Sometimes it only provides partial information. If you check the paper more closely yourself, you’ll find that tools like ChatGPT don’t capture the truly important information.''
\end{quote}

Participants further raised these concerns for junior researchers. As reflected in P5, they said ``Students would never know the applicability of information.'' They explained that when individuals are new to a field and lack sufficient exposure to the depth of prior work that defines a meaningful and novel contribution, using AI may obscure gaps in understanding rather than reveal them. This concern was further reflected in P6's account of abandoning AI tools after repeated experiences of receiving confidently stated but poorly grounded responses: ``No matter what question you ask, the platform always gives you an answer, even when it's not directly related.''

Taken together, these experiences reveal that the assertive tone of AI outputs not only creates a false sense of reliability but also imposes an additional accountability burden: one driven less by researchers' own intellectual interests and more by the responsibility of maintaining scholarly standards.

\subsubsection{Platform-level opacity around data governance.} 
While much of the discourse on accountability focuses on the validity of AI-generated results, participants (2 of the 15) also identified a secondary, equally important issue of protecting their own data. There is concern about the lack of accountability in how AI platforms handle researchers' prompts containing valuable, unpublished ideas. P11 highlighted the risk of this unintended disclosure:
\begin{quote}
    ``They [AI platforms] will leverage the prompt you share for training, which has the potential to leak your research question or research data.''
\end{quote}
The lack of clear data governance worsens this concern. The specific forums (as defined in the accountability section of Section \ref{subsec:RAI in AI4Science}) to which AI companies must respond are not visible to end users. P12 noted:
\begin{quote}
    ``Not knowing how much of my personal data is being stored, where it is being stored, and who has access to it, and also the limitation of the system.''
\end{quote}
These concerns reveal that accountability in AI-mediated research is not solely a question of output validity but also of institutional answerability: the degree to which AI platforms can be held responsible for the researchers' data they collect, store, and potentially repurpose.

\subsubsection{Human retains ultimate scholarly judgment.}
Participants (9 of 15 participants) consistently noted AI as unable to bear responsibility for core scholarly judgment, particularly around value, relevance, and correctness. While AI tools could confidently generate answers, participants emphasized that confidence was decoupled from validity, thereby placing ultimate accountability on researchers. For example, P5 said: 
\begin{quote}
    ``I don't rely on AI for formulating research questions or hypotheses, or finding sources to back up my arguments. So mostly like, I use AI tools for very trivial, minor things that can support my arguments rather than formulating my arguments.''
\end{quote}
This pattern infers a broader strategy of preserving human agency over the scholarly judgments that define research contributions, while selectively delegating lower-stakes preparatory tasks to AI.

\subsubsection{Restrict AI to organizational and peripheral tasks.}
Rather than rejecting AI tools entirely, participants (10 of 15) found value in using them for tasks that support the research process without directly influencing schoarly judgment. These included helping them track decision points during literature reviews, summarize large bodies of work, or document intermediate reasoning steps that might otherwise be lost. For example, P4 said:
\begin{quote}
    ``What I'm hoping will be useful is that the tool has more note-taking capabilities, and has folders or spaces where I can organize papers I think I'll return to, and [that] it can remind me of the aim of a given paper, just to get a general understanding of it.''
\end{quote}
This strategy allowed participants to derive efficiency gains from AI tools while maintaining clear boundaries around the scholarly judgments they considered non-delegable, preserving accountability by keeping the intellectual core of research firmly in human hands.

\subsection{Transparency}
Participants reported that the transparency challenge stems from both the system-level opacity of information retrieval and the logical-level opacity of narrative construction. They also noted that hallucinations exemplify this transparency problem. To manage these issues in real time, participants had to rely on heuristic-based strategies that were often unreliable, as well as additional manual verification and constrained prompting to request more accurate information.
 
\subsubsection{Blackbox opacity across retrieval and narrative construction.}
The primary transparency barrier is the ``blackbox'' nature of both the system-level and logical-level opacity. On one hand, they (9 of 15 participants) are concerned about where the data comes from, such as unclear sources and retrieval pipelines, unknown training, and database coverage. For example, P12 said the lack of information regarding the database and coverage limits the tool's utility for rigorous research reports:
\begin{quote}
    ``I don't even know how many databases it has access to. The black box nature of it kind of makes it a limitation for some studies, because you cannot report with certainty the sources or the data.''
\end{quote}
On the other hand, they are concerned about how AI decides what is relevant and how it formulates content. P6 highlighted the intractability of AI curation logic, noting that researchers cannot determine whether an omission reflects a deliberate selection or a random failure of the system:
\begin{quote}
    ``What I don’t like is that I don’t understand the logic behind it. For example, when it adds papers after a search, I don’t know how that curation works, based on the relevance or just based on what?''
\end{quote}
Without traceability at both levels, even when citations or summaries are included in AI outputs, the fundamental scientific requirement for provenance is unmet, making it impossible for participants to adopt them. 

\subsubsection{Hallucinations as transparency failures.}
While hallucination is often treated as a technical problem in generative AI, participants (7 of 15) in our study framed it as a profound failure of transparency. The primary issue is not just the inaccuracy itself, but the way the fluent, assertive form of the output masks a lack of underlying substance. 

Unlike traditional retrieval systems that present sources for direct inspection, LLMs perform a ``synthetic synthesis'' that often abstractly condenses or shallowly connects information, losing the concrete grounding of the original text. P5 directly pointed out:
\begin{quote}
    ``This is absolutely not a correct citation...[the] abstract should show a more concrete summary rather than bringing sentences from here and there.''
\end{quote}
This ``shallowing'' of content manifests in two problematic scenarios that undermine the provenance of research. First, attribution displacement occurs when information is synthesized correctly but misattributed to an incorrect source or presented without credit entirely. Second, synthetic blending involves the seamless integration of factual inaccuracies alongside valid citations, making it nearly impossible to distinguish verified scholarship from synthetic fabrication without exhaustive manual effort. Consequently, both scenarios led to a user pattern of redundant verification, where some participants treated AI-generated claims with suspicion and asked, as P10 noted:
\begin{quote}
    ``I [was] just asking the references for those things, because, as you know, there is a general concern of authenticity in the ChatGPT; it just creates dummy information and all.''
\end{quote}

\subsubsection{Social credibility heuristics.}
In response to the structural opacity of AI tools, researchers (15 of 15) developed several mitigation strategies. However, these workarounds are often fragile, as they heavily rely on a researcher pre-existing heuristics and expertise, and sometimes cost a significant expenditure of time. This complicates the intended efficiency of AI tools.

For example, personal familiarity with an author's prior work appeared to provide participants with implicit meta-information regarding the credibility and validity of new research. If Author A had previously produced relevant and high-quality work, participants were more inclined to trust the substance of Paper Y authored by the same individual, even when it appeared in a different venue or with unfamiliar co-authors. P9 said:
\begin{quote}
    ``I'm still well aware of my research topic...I know these authors are key figures in this research area...they are making an important note in this research network.''
\end{quote}
In contrast, when such familiarity was absent, participants relied more heavily on other indicators, such as methodological rigor, citation counts, topical relevance, and publication venue.

\subsubsection{Redundant manual verification.}
However, when the above social heuristics were insufficient, participants \hl{(8 of 15)} defaulted to a pattern of redundant, manual verification. This involved repeatedly verifying names, publication years, and citations, often re-checking the same information multiple times. For example, P13 reported:
\begin{quote}
    ``I do prefer reading the abstract first... just to make sure that it's not hallucinating or just completely bringing things out of context because it did happen.''
\end{quote}
This behavior reflects a broader lack of trust in AI-generated data, even when systems employ retrieval-based generation techniques (we further elaborate on this reliance pattern in Section \ref{findings:reliance}).

\subsubsection{Constrained prompting.}
Another commonly used strategy is more detailed or constrained prompting. As P11 said: ``sometimes you need to write a long paragraph in the prompt, otherwise it sounds relevant but not specific.'' However, participants \hl{(3 of 15)} also noted that this shifts additional effort and time burdens onto researchers. Moreover, for newer students entering emerging research areas, crafting precise and exhaustive prompts may be a significant challenge in itself. As a result, transparency issues not only persist but also exacerbate existing inequalities in expertise and experience.

\subsection{Trust} \label{findings:reliance}
Trust emerged as a cross-cutting dimension that shaped whether and how participants chose to rely on AI tools across all stages of their research workflows. Tensions centered on the fragility and context-dependence of AI trust, and the homogenization of outputs that undermined participant confidence in the intellectual depth of AI outputs. In response, participants developed two compensatory strategies: building trust progressively through logical auditing and hedging AI use for low-stakes tasks.
 
\subsubsection{A fragile, context-dependent, and continuously re-established process.}
Rather than developing stable, generalized trust in AI tools over time, participants 11 of 15) found that trust had to be actively constructed and continuously re-established through acts of verification. A key reason for this fragility is that AI tools often appear to perform reliably without actually being so, placing participants in a position of systematic suspicion, where even outputs that seem well-grounded cannot be trusted without independent verification. As P7 observed:
\begin{quote}
    ``If it says something wrong, and you're able to point it out... I specifically remember telling ChatGPT that the citation you just gave me doesn't exist. And then it goes into that conversational apology mode... it'll throw in citations and references just to make sure that it gives the impression that it has a lot of capacity — kind of like, fake it till you make it.''
\end{quote}
P7 further noted that irrelevant citations were not always factually fabricated but simply disconnected from the research context: ``I'm sure that's a real citation, but there really is no connection.'' This distinction is important: trust fragility in AI-mediated research is not solely a hallucination problem but a provenance problem; the issue is not only whether a source exists, but whether it belongs.

This trust-building process was grounded in experiential benchmarking, where participants used their domain expertise as a baseline to evaluate new information. For example, P15 described assessing the AI output by comparing the statistical rigor of plots and error bars against what they have done in other projects:
\begin{quote}
    ``I start by thinking about whether these fits seem better, worse, or about the same as the ones I’ve worked with in other projects. What I pay attention to are things like the number of points in each plot and the error bars. In fact, the plot with more data points looks quite good to me, compared to what I’ve seen in my other projects.''
\end{quote}
This pattern reveals that trust in AI tools was calibrated through a gut-check against personal professional history rather than derived from any intrinsic property of the tool itself.

\subsubsection{Output homogenization undermines perceived intellectual depth.} A second trust tension emerged not from inaccuracy but from the perceived shallowness and generic quality of AI-generated outputs. When AI tools presented research ideas or summaries in overly broad terms, participants (7 of 15) lost confidence in the tool's capacity to support genuine scholarly contributions. As P14 observed:
\begin{quote}
    ``The summary is pretty basic. I basically know all of it... But what I'm more curious about is how changing or tweaking these elements will result in different outcomes...'' 
\end{quote}
This experience of homogenization, where AI outputs flatten the nuance and specificity that distinguish meaningful scholarly contributions, reflects a deeper trust concern: that AI tools may be capable of aggregating existing knowledge but not of supporting the kind of original, contextually grounded thinking that defines research at its best. For participants, this perceived intellectual shallowness made it difficult to calibrate how much trust to place in AI tools even when outputs were factually accurate, underscoring that trust in AI-mediated research is as much about intellectual depth as it is about correctness.

\subsubsection{Progressive trust-building through logical auditing.} Rather than accepting outputs at face value, participants like P10 actively interrogated the tool ``internal logic'' by cross-referencing AI-generated claims with established facts and requesting the system's underlying reasoning. As P10 noted, ``I am trying to build the trust here by seeing how the system is actually working.'' Trust, in this sense, was earned incrementally through repeated verification rather than assumed from the outset. 

\subsubsection{Hedging AI use to low-stakes tasks.}
Rather than treating AI as an authoritative source, participants (10 of 15) used strategic hedging by restricting AI use to low-stakes tasks while leaving the core scientific activities on their own. For example, P4 described the use of AI for broad context understanding: 
\begin{quote}
    ``I have used tools such as ChatGPT, sometimes Perplexity... and that was generally having a background study, a general overview of literature, or finding a research gap.''
\end{quote}
The participant then limited the use of AI to initial, surface-level exploration and transitioned to conducting deeper analysis independently:
\begin{quote}
    ``I pay attention a little bit to the method that they were using, and take notes like... And then once I'm looking at the later figures, I'm starting to think I should read some of the paper itself.''
\end{quote}
This transition from AI-assisted work to manual verification reflects a concern over the flattening effect inherent in AI-generated summaries. During the literature synthesis stage, P13 cautiously mentioned the use of AI because the synthesis could easily omit what ``is really important,'' to underscore scholar's unique academic standpoint:
\begin{quote}
    ``One reason I hesitate to use generative AI for literature reviews is that I don’t see a literature review as just a report, listing what one paper did and another did, but as a way to contextualize prior work and explain how my research builds on it.''
\end{quote}
Overall, participants described an ongoing tension between skepticism and reliance. While AI tools offer efficiency and accessibility, their limitations in transparency, originality, and contextual understanding make it difficult for researchers to confidently calibrate how much trust to place in them. This tension underscores the need for clearer norms, interfaces, and accountability structures to support informed and responsible use of AI in scientific research.

\section{Discussion}

The advent of AI has catalyzed a fundamental shift \citep{xu2021artificial, wang2023scientific} in which AI-driven approaches accelerate scientific discovery by moving beyond human-led analysis toward data-driven inquiry \citep{humphreys2020automated, Wang2023-hl}. This integration is not merely a technical trend but a structural evolution in the scientific community, driven by market forces and increasingly formalized within scholarly discourse \citep{cruz2025epistemic, traberg2026ai}. We argue that advancing AI-mediated research cannot be reduced to improving models alone and it requires treating AI as part of a collective sociotechnical system in which transparency, accountability, and trust are actively designed for and socially negotiated. While theoretical frameworks define the ideals of RAI \citep{schiff2020principles}, a growing body of empirical research documents how these principles manifest, or fail to, in real-world scientific practice \citep{bach2025insights, reuel2025responsible}. Survey work captures the breadth of this concern: a study of over 1,600 researchers worldwide found that more than 50\% expressed heightened anxieties around over-reliance, entrenched bias, and AI-facilitated research fraud \citep{van2023ai}. 

Our think-aloud findings extend this discourse by documenting how these concerns are enacted in practice across our three focal constructs. Regarding accountability, participants found that the confident tone of AI outputs obscured the epistemic uncertainty inherent in scholarly judgment, leading them to restrict AI use on core evaluative tasks such as assessing literature relevance and correctness. Regarding transparency, the opacity of AI retrieval and generation processes made establishing provenance impossible, prompting participants to rely on social credibility heuristics, manual verification, and constrained prompting. Regarding trust, participant confidence in AI was fragile and context-dependent, built incrementally through logical auditing and deliberately bounded to lower-stakes tasks.

Taken together, participant experiences reveal shared challenges that reflect not merely usability failures but a deeper restructuring of epistemic accountability within scientific practice \citep{clark2025epistemic}. While AI tools offer genuine efficiency gains, our findings suggest these are consistently accompanied by normative tensions that existing RAI frameworks have not yet adequately addressed. Current AI development for science disproportionately emphasizes tasks where large language models are most effective \citep{akbar2025use, khalifa2024using}, yet this emphasis risks conflating information aggregation with genuine discovery, and efficiency with epistemic rigor.

These negotiations are significant precisely because they are happening in real time, without institutional scaffolding or shared norms to guide them. Current AI tools redistribute workload asymmetrically, reducing researcher agency in ways that are not always visible or intentional \cite{kapania2025m}, while the AI-as-partner paradigm increasingly encouraged in the literature surfaces unresolved questions about human-AI agency and shifting roles in knowledge production \cite{phan2025ai}. \hl{Participants worried that such tools may accelerate paper production and early-stage ideation while quietly eroding the practices of verification, theoretical grounding, and experimental rigor that underpin scientific credibility }\cite{khalifa2024using}. \hl{Below, we outline alternative trajectories actionable at both individual and institutional levels. While the challenges associated with these technologies are novel, they are not unprecedented. Prior transitions in scholarly infrastructure, such as the introduction of retrieval-based literature search, offer instructive precedents for deliberate, measured adoption over rapid or uncontrolled deployment. We then turn to the central issue of trust, arguing that there is no one-size-fits-all approach to establishing or sustaining it within AI-supported research workflows, and that meaningful progress requires attending to the situated, discipline-specific norms through which trust is actually built and broken in scientific communities.}

\subsection{Phased Transition to New Tools}

Given the uncertain nature of these tools and concerns around trust, transparency, and accountability, it is essential (both in design and in principle) to adopt a controlled rollout. From a design perspective, one way to achieve this is to emphasize not only performance, but also the \textbf{retention of essential functionalities and workflow support from earlier platforms}. Rather than simply resisting new interfaces, participants described situations in which key features present in established tools (such as timeline filtering, author filtering, or previewing search results in \textit{Google Scholar} and \textit{Semantic Scholar}) were missing in newer platforms like Research Rabbit or Elicit AI. These omissions often forced researchers to combine multiple tools or devise workarounds to accomplish tasks that were previously straightforward, highlighting gaps in workflow continuity rather than reluctance to learn.

Participants noted that functionalities such as the ability to store, synchronize, and share literature libraries (e.g., via Zotero or Mendeley integration) were critical for supporting research efficiency and early adoption. However, some newer AI-focused tools, including Elicit AI, offered limited support for such integrations, creating friction in practical usage. Our findings suggest that \textbf{preserving core functionalities and supporting seamless workflow continuity}, rather than simply creating novel features, can significantly enhance tool adoption and effectiveness. 

\subsection{Channels for Trust Establishment}
 
Ethical concerns have been raised about AI’s role in authorship and attribution when AI tools are used in research ideation and writing \cite{korinek2023language, stokel2023chatgpt}. Our findings suggest that researchers address these concerns through ad hoc practices. Prior work similarly reports trust and over-reliance issues in MetaWriter, an AI system supporting peer reviewers in writing meta-reviews \cite{sun2024metawriter}, with mistrust exacerbated by limited transparency into the system’s decision-making processes and intermediate outputs. 

Efforts centered on the verification and validation of AI-generated outputs, as participants emphasized the critical need for mechanisms that ensure reliability. Participants often engaged in innovative practices to circumvent the issues of reliability. These range from manually verifying to prompting LLMs by embedding paper metadata. \textit{Elicit AI} was frequently cited as a promising example of how explainable practices could be embedded more systematically into tool design. Given that research is a precise endeavor with little tolerance for error, hallucinated content, or non-factual claims, participants underscored that inaccuracies not only \textbf{undermine researcher credibility} but also \textbf{risk propagating misinformation across subsequent work}. In our study, participants consistently cross-checked, verified, and validated each AI-generated output, reflecting a broader demand for accountability in these systems. Future tools, therefore, must invest significantly in features that support verifiability and transparency. This concern is particularly salient for early-career researchers, who may have limited domain expertise and thus be more susceptible to being misled by erroneous outputs. In such cases, the difficulty of independently assessing veracity further amplifies the importance of tools that embed strong mechanisms for validation, thereby fostering trust and accountability in the research process. Through this work, there also emerged relevant design ideas centering the role of agency in verification and support of qualifying information around novel features. 

In particular, participants underscored the value of metadata, which they regarded as critical for informing decision-making processes and assessing the relevance of retrieved materials. Their reflections revealed a set of nuanced expectations for how content should be structured and presented in order to be meaningfully integrated into research workflows. For instance, in a review of human-AI teaming research,\citep{berretta2023defining} observed that there was a common theme of researchers pushing for more explainable systems to help make decisions. It emerged that tools like Elicit AI are moving in the right direction towards establishing trust by including verification pipelines within their design.

\subsection{Limited Scoping of Use}

A central focus of recent work has been the acceleration of research workflows. Our findings indicate that AI tools substantially contribute to expediting these processes. Participants reported that such tools facilitated faster discovery of \textbf{relevant literature, efficient parsing of research papers, and access to concise starting-point summaries}. Notably, AI tools appeared to be particularly valuable during \textbf{the preliminary exploration} of unfamiliar research domains, where participants tended to rely on them as an initial point of entry. AI tools were perceived as helpful in several dimensions of the research process. Participants emphasized their value in enabling faster discovery of relevant literature and in casting a wider net when engaging with new or emerging research topics. In addition, these tools supported the identification of specific locations within texts where answers to targeted questions might be found, thereby streamlining information retrieval. While adoption for research ideation remains partial and characterized by some resistance, participants nonetheless reported deriving significant benefit from the use of AI tools in language-related tasks, particularly for writing and rewriting academic text.

With the range of terms associated with the role that AI plays (assistants, copilots, consultants, and so on) there were no researchers who relied on AI for end-to-end work \cite{chen2025generative}. The need for human-in-the-loop intervention, to put it mildly, reflects \textbf{the level of control researchers wished to maintain} throughout their workflows. The speed, however, comes from the ability of AI tools to quickly summarize and strip text of extraneous or overly elaborate language rather than replace substantive intellectual contributions. In practice, researchers reported using AI to accelerate preliminary steps such as scanning a large set of papers, extracting key insights, and generating concise overviews. This acceleration was viewed as particularly valuable in the early exploratory phases of research, where efficiency in identifying relevant sources and constructing an initial understanding of a new topic can have significant downstream benefits. Yet, participants emphasized that this efficiency did not negate the necessity of careful reading, critical evaluation, and contextual integration—tasks that they firmly reserved for themselves. This reflects a documented tendency for individuals to retain responsibility for cognitively demanding and critical thinking tasks, while increasingly offloading preparatory or data-oriented tasks onto AI-based tools. 

 \section{Limitations and Future Work}

\hl{Our findings are subject to several important limitations. Our participant pool draws primarily from academic researchers, excluding industry contexts where epistemic norms and accountability structures around AI may differ substantially. The two tools selected, Research Rabbit and Elicit, also represent a snapshot of a rapidly evolving landscape; both have undergone updates since data collection, and our findings should be understood as documenting researcher behavior under specific technological conditions rather than as stable characterizations of either system. Additionally, think-aloud methodology, while well-suited for capturing real-time reasoning, provides only a partial window into researcher cognition, as verbalization is effortful and incomplete, and the act of articulating judgment may itself alter the processes under observation. Participant interactions with AI tools during the study were necessarily shaped by the task structure we provided, meaning that the prompts they issued and the outputs they encountered were not fully representative of the open-ended, self-directed AI use characteristic of naturalistic research practice. Finally, our findings are sensitive to the specific tasks and prompts participants encountered; reasoning and judgments may have differed under alternative task framings or varying levels of tool familiarity. Taken together, these constraints point toward the need for longitudinal, naturalistic studies that can track how researchers' epistemic negotiations with AI tools evolve as both the tools and the norms surrounding them mature.} 

Future work could build on these findings by \textbf{investigating domain-specific requirements}, as research practices and workflows vary widely across disciplines. Larger and more diverse participant pools would further enhance understanding of how AI tools are adopted and adapted in different contexts. Additionally, integrating \textbf{behavioral logs, interaction data, and quantitative metrics} could provide more objective measures of tool efficacy, capturing not only subjective reflections but also patterns of use, productivity impacts, and workflow changes. Combining qualitative insights with such quantitative analyses would enable a more comprehensive evaluation of AI tools, supporting the design of systems that are both \textbf{flexible across domains and effective in enhancing research outcomes}.

\section{Conclusion}

In this work, we emphasize the importance of mediating judgment over AI-supported and AI-generated outputs through a Responsible AI (RAI) lens. Drawing on empirical evidence captured during task performance rather than retrospective reflection, we surface the concerns researchers experience when engaging with AI-mediated research, as well as the strategies they currently employ to address them. Together, these findings motivate the need for a slow, deliberate, and well-regulated adoption of AI tools in scientific practice to safeguard scientific integrity.

\section*{Acknowledgments}

Our research would not have been possible without our participants, and we thank our anonymous reviewers for their valuable feedback. This work was supported in part by the NSF-Simons AI Institute for Cosmic Origins (CosmicAI)\footnote{\url{https://www.cosmicai.org/}} through NSF Cooperative Agreement 2421782) and the Simons Foundation (grant MPS-AI-00010515), as well as by Good Systems (a UT Austin Grand Challenge dedicated to developing responsible AI technologies)\footnote{\url{https://goodsystems.utexas.edu/}}, the School of Information, and a Research Fellowship from UT Austin. The statements made herein are solely those of the authors alone and do not necessarily reflect the positions of the supporting institutions.

\section*{Generative AI Disclosure Statement}

We used Generative AI (Claude 4.5 and Gemini 3.0) for generic conversational brainstorming, followed by entirely manual paper writing. Upon completion of the manuscript, we used Grammarly AI or ChatGPT for grammar corrections and sentence restructuring and an Agentic Reviewer (\url{https://paperreview.ai/}) to iteratively gather feedback and improve the paper. 

\bibliographystyle{ACM-Reference-Format}
\bibliography{reference}

\end{document}